\documentclass[conference]{IEEEtran}
\usepackage{amsmath}
\usepackage{amssymb}
\usepackage[dvips]{graphicx}
\usepackage{epsfig}

\newcommand{\z}{\mathbf{z}}
\newcommand{\h}{\mathbf{h}}

\newcommand{\hh}{\mathbf{H}}
\newcommand{\w}{\mathbf{w}}

\newcommand{\I}{\mathbf{I}}

\newcommand{\bv}{\mathbf{v}}

\newcommand{\bu}{\mathbf{u}}

\newtheorem{lemma:bitenergylow}{Lemma}
\newtheorem{lemma:bitenergyhigh}[lemma:bitenergylow]{Lemma}

\newtheorem{prop:asympcap}{Theorem}
\newtheorem{prop:flashminbitenergy}[prop:asympcap]{Theorem}
\newtheorem{prop:flashbitenergy}[prop:asympcap]{Theorem}
\newtheorem{prop:pasympcap}[prop:asympcap]{Theorem}

\begin{document}


\title{Collaborative Relay Beamforming for Secure Broadcasting}

%\author{\authorblockN{Michael Shell} \and
%\and \authorblockN{James Kirk\\ and Montgomery Scott}
%\authorblockA{Starfleet Academy\\
%San Francisco, California 96678-2391\\ Telephone: (800)
%555--1212\\ Fax: (888) 555--1212}}

% avoiding spaces at the end of the author lines is not a problem with
% conference papers because we don't use \thanks or \IEEEmembership

\author{\authorblockN{Junwei Zhang and Mustafa Cenk Gursoy}
\authorblockA{Department of Electrical Engineering\\
University of Nebraska-Lincoln, Lincoln, NE 68588\\ Email:
junwei.zhang@huskers.unl.edu, gursoy@engr.unl.edu}}

\maketitle
\begin{abstract}\footnote{This work was supported by the National Science Foundation under Grant CCF -- 0546384 (CAREER).}
In this paper, collaborative use of relays to form a beamforming
system with the aid of perfect channel state information (CSI) and
to provide communication in physical-layer security between a
transmitter and two receivers is investigated. In particular, we
describe decode-and-forward based null space beamforming schemes and
optimize the relay weights jointly to obtain the largest secrecy
rate region. Furthermore, the optimality of the proposed schemes is
investigated by comparing them with the outer bound  secrecy rate region.
\end{abstract}

\section{introduction}
The open nature of wireless communications allows for the signals to
be received by all users within the communication range. Thus,
secure transmission of confidential messages is a critical issue in
wireless communications. This problem was first studied in
\cite{wyner} where Wyner identified the rate-equivocation region and
established the secrecy capacity of the discrete memoryless wiretap
channel in which eavesdropper's channel is a degraded version of the
main channel. Later, Wyner's result was extended to the Gaussian
channel in \cite{cheong} and recently to fading channels in
\cite{Gopala}. In addition to the single antenna case, secure
transmission in multi-antenna models is addressed in \cite{shafiee}
-- \cite{khisti}. For multi-user channels, Liu \emph{et al.}
\cite{Liu} presented inner and outer bounds on secrecy capacity
regions for broadcast and interference channels. The secrecy
capacity of multi-antenna broadcasting channel is obtained in
\cite{Liu1}. Moreover, it's well known that that users can cooperate to form a
distributed multi-antenna system by relaying. Cooperative relaying
with secrecy constraints was recently discussed in \cite{dong}--\cite{dong1} .

In this paper, we study the relay-aided secure broadcasting
scenario. We assume that the source has two independent messages, each of which is
intended for one of the receivers but needs to be kept
asymptotically perfectly secret from the other. This is achieved via
relay node cooperation in decode and forward fashion to produce
virtual beam points to two receivers. The problem is formulated as a problem of designing the
relay node weights in order to maximize the secrecy rate for both receivers
for a fixed total relay power. We assume that the global channel state information (CSI) is available
for weight design. Due to the difficulty of the general optimization
problem,  we propose null space beamforming transmission schemes and
compare their performance with the outer bound secrecy rate region.

\section{Channel}
\begin{figure}
\begin{center}
\includegraphics[width = 0.5\textwidth]{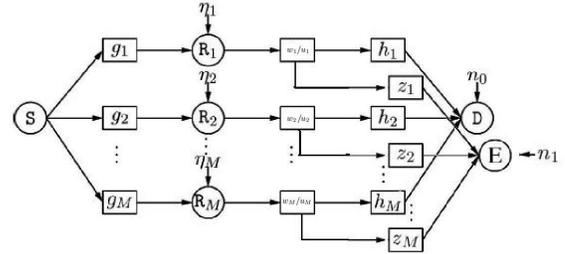}
\caption{Channel Model1} \label{fig:channel}
\end{center}
\end{figure}

We consider a communication channel with a source $S$, two
destination nodes $D$ and $E$, and $M$ relays $\{R_m\}_{m=1}^M$ as depicted in
Figure \ref{fig:channel}. We assume that there is no direct link
between $S$ and $D$, and $S$ and $E$. We also assume that relays work synchronously
and multiply the signals to be transmitted by complex weights to produce virtual beam points
to $D$ and $E$. We denote the channel fading coefficient between $S$ and $R_m$ as
$g_m\in \mathbb{C}$ , the channel fading coefficient between $R_m$ and $D$ as $h_m\in
\mathbb{C}$, and the channel coefficient between $R_m$ and $E$ as $z_m\in
\mathbb{C}$. In this model, the source $S$ tries to transmit confidential messages to
$D$ and $E $ with the help of the relays .
It is  obvious that our channel is a two hop relay network. In the
first hop, the source $S$ transmits $x_s$ which contains the confidential
messages intended for both $D$ and $E$ to the relays with 
power $E[|x_s|^2]=P_s$. The received signal at relay $R_m$ is
given by
\begin{align}
y_{r,m}=g_m x_s+\eta_m
\end{align}
where $\eta_m$ is the background noise that has a Gaussian
distribution with zero mean and a variance of $N_m$.

In the first hop, the secrecy rates for destination $D$  and $E$ lie in the following triangle region.
\begin{align}\label{sr}
 &R_d\geq 0 \,\, \text{and} \,\, R_e\geq 0 \\
 &R_d+R_e \leq \min_{m=1,\ldots,M} \log\left(1+\frac{|g_m|^2P_s}{N_m}\right)
\end{align}
where $R_d$ and $R_e$ denote the secrecy rates for destination $D$
and $E$, respectively.

\section {Relay Beamforming} 

We consider the scenario in which relays are much more closer to the source
than the destinations, and hence, the first-hop rate does not become a
bottleneck of the whole system. Due to this assumption, we in the following
focus on characterizing the secrecy rate region of the second-hop. We
consider the decode-and-forward relaying protocol in which each
relay $R_m$ first decodes the message $x_s$,and subsequently scales the decoded messages to obtain $x_r=w_m x_d + u_m x_e$, where $w_m$ and $u_m$ are the weight values.
$x_d$ and $x_e$ are independent, zero-mean, unit-variance Gaussian signals which include the confidential messages to $D$
and $E$, respectively. Under these assumptions, the output power of relay  $R_m$ is
\begin{align}
E[|x_r|^2]=E[|w_m  x_d + u_m x_e|^2]=|w_m|^2+|u_m|^2
\end{align}
The received signals at the destination nodes $D$ and $E$ are the
superpositions of the signals transmitted from the relays. These
signals can be expressed, respectively, as
\begin{align}\label{trr}
y_d&=\sum_{m=1}^M h_m w_m  x_d + \sum_{m=1}^M h_m u_m  x_e+n_0\nonumber \\
&=\h^\dagger \w x_d + \h^\dagger \bu x_e +n_0 \\
 y_e&=\sum_{m=1}^M z_m w_m  x_d +\sum_{m=1}^M z_m  u_m  x_e +n_1 \nonumber \\
&=\mathbf{z}^\dagger \w x_d +\mathbf{z}^\dagger  \bu x_e +n_1 \label{trr_2}
\end{align}
where $n_0$ and $n_1$ are the  Gaussian background noise  components at
$D$ and $E$, respectively, with zero mean and variance $N_0$.
Additionally, we have above defined $\mathbf{h}=[h_1^*,....h_M^*]^T, \mathbf{z}=[z_1^*,....z_M^*]^T$,
$\w=[w_1,...w_M]^T$,  and $\bu=[u_1,...u_M]^T$.  In these notations, while superscript
$*$ denotes the conjugate operation, $(\cdot)^T$ and  $(\cdot)^\dagger$ denote the
transpose  and conjugate transpose , respectively, of a matrix or
vector. From the transmitting and receiving relationship in
(\ref{trr}) and (\ref{trr_2}),  we can see that the channel we consider can be treated as an interference channel with secrecy constraints
studied in \cite{Liu}. The achievable secrecy rate region  is shown
to be
\begin{align}
0 \leq R_d \leq &\log \left(1+\frac{|\sum_{m=1}^M h_m
w_m|^2}{N_0+|\sum_{m=1}^M
h_m u_m|^2}\right) \nonumber
\\
&- \log \left( 1+\frac{|\sum_{m=1}^M z_m w_m|^2}{N_0}\right)\label{Rd}
\\
0 \leq R_e \leq &\log \left(1+\frac{|\sum_{m=1}^M z_m
u_m|^2}{N_0+|\sum_{m=1}^M z_m w_m|^2}\right)\nonumber
\\
&- \log \left(
1+\frac{|\sum_{m=1}^M h_m u_m|^2}{N_0}\right).\label{Re}
\end{align} 
In this paper, we address the joint optimization $\{w_m\}$ and
$\{u_m\}$ with the aid of perfect CSI, and hence identify
the optimal collaborative relay beamforming (CRB) direction that maximizes the
secrecy rate region given by (\ref{Rd}) and (\ref{Re}). Since the optimization
problem above is in general intractable, we investigate suboptimal schemes.

\subsection{Single Null Space Beamforming} \label{subsec:singlenull}

In this scheme, we choose one user's (e.g., $E$) beamforming vector
(e.g., $\bu$) to lie in the null space of the other user's channel.
With this assumption, we eliminate the user $E$'s interference on
$D$ and hence $D$'s capability of eavesdropping on $E$. Mathematically, this
is equivalent to  $|\sum_{m=1}^M h_m u_m|^2=\h^\dagger \bu=0$, which
means $\bu$ is in the null space of $\h^\dagger$.

%Now the channel simplify to
%\begin{align}
%y_d&=\h^\dagger \w x_d  +n_0 \\
% y_e&=\mathbf{z}^\dagger \w x_d +\mathbf{z}^\dagger \bu x_e
%+n_1
%\end{align}
We further assume  $\alpha$ fraction of total relay transmitting
power $P_r$ is used for sending confidential message to $D$. Under these assumptions, we can solve the optimization problem in
(\ref{Rd}). The maximum $R_d$ can be computed as
\begin{align}\label{dftotal}
&R_{d,m}(\h,\z,P_r,\alpha)\nonumber\\
&=\max_{\w^\dagger \w\leq \alpha P_r}\log\frac{N_0+|\sum_{m=1}^M
h_m w_m|^2}{N_0+|\sum_{m=1}^M z_m
w_m|^2}\\
&=\log  \max_{\w^\dagger \w \leq \alpha P_R}\frac{N_0+|\sum_{m=1}^M
h_m
w_m|^2}{N_0+|\sum_{m=1}^M z_m w_m|^2}\\
&=\log  \max \frac{\w^\dagger(\frac{N_0}{\alpha P_r}\I+\h
\h^\dagger)\w}{\w^\dagger(\frac{N_0}{\alpha P_r}\I+\z \z^\dagger)\w}\\
&=\log  \max \frac{\w^\dagger(N_0\I+\alpha P_r\h
\h^\dagger)\w}{\w^\dagger(N_0 \I+ \alpha P_r\z \z^\dagger)\w} \label{ra1}\\
&=\log \lambda_{max}(N_0\I+\alpha P_r\h \h^\dagger,N_0\I+\alpha
P_r\z\z^\dagger)\label{ra}
\end{align}
Here, we use the fact that (\ref{ra1}) is the Rayleigh quotient problem,
and its maximum value is as given in (\ref{ra}) where
$\lambda_{max}(\mathbf{A},\mathbf{B})$ is the largest generalized
eigenvalue of the matrix pair $(\mathbf{A},\mathbf{B})$. Note that we
will also use $\lambda_{max}(\cdot)$ to denote largest eigenvalue of
the matrix in later discussion. The optimum beamforming weights $\w$
is
\begin{align}\label{woptdfto}
\w_{opt}=\varsigma\psi_w
\end{align}
where $\psi_w$ is the eigenvector that corresponds to
$\lambda_{max}(N_0\I+\alpha P_r\h \h^\dagger,N_0\I+\alpha
P_r\z\z^\dagger)$ and  $\varsigma$ is chosen to ensure
$\w_{opt}^\dagger \w_{opt} = \alpha P_r$.

Now we turn our attention to the  maximization of $R_e$ when $\w = \w_{opt}$. Note that
$N_0+|\sum_{m=1}^M z_m w_m|^2$ is a constant denoted by $N_t$, Due
to the null space constraint, we can write $\bu=\hh_{h}^\bot \bv$,
where $\hh_{h}^\bot$ denotes the projection matrix onto the null
space of $\h^\dagger$. Specifically, the columns of $\hh_{h}^\bot$ are
orthonormal vectors which form the basis of the null space of
$\h^\dagger$. In our case,  $\hh_{h}^\bot$ is an $M\times(M-1)$
matrix. The power constraint $\bu^\dagger \bu= \bv^\dagger
{\hh_{h}^\bot}^\dagger \hh_{h}^\bot \bv=\bv^\dagger \bv\leq
(1-\alpha)P_r$.

The maximum $R_e$ under this condition can be computed as
%\begin{align}\label{dftotal}
%&R_{e,m}(\h,\z,P_r,\alpha)\nonumber \\
%&=\max_{\bu^\dagger \bu\leq (1-\alpha) P_r}\log
%\left(1+\frac{|\sum_{m=1}^M z_m u_m|^2}{N_t}\right)\\
%&=\log\max_{\bu^\dagger \bu\leq (1-\alpha) P_r} \left(
%\frac{\bu^\dagger(\frac{N_t}{(1-\alpha)P_r}\I +\z \z^\dagger)
%\bu}{\bu^\dagger \frac{N_t}{(1-\alpha)P_r}\I \bu}\right)\\
%&=\log\max_{\bv^\dagger \bv \leq (1-\alpha) P_r} \left(
%\frac{\bv^\dagger {\hh_{h}^\bot}^\dagger(\frac{N_t}{(1-\alpha)P_r}\I
%+\z \z^\dagger) \hh_{h}^\bot\bv}{\bv^\dagger {\hh_{h}^\bot}^\dagger
%\frac{N_t}{(1-\alpha)P_r}\I \hh_{h}^\bot \bv}\right)\\
%&=\log
%\lambda_{max}({\hh_{h}^\bot}^\dagger(\frac{N_t}{(1-\alpha)P_r}\I +\z
%\z^\dagger) \hh_{h}^\bot,{\hh_{h}^\bot}^\dagger
%\frac{N_t}{(1-\alpha)P_r}\I \hh_{h}^\bot )
%\end{align}
\begin{align}\label{dftotal}
&R_{e,m}(\h,\z,P_r,\alpha)\nonumber \\
&=\max_{\bu^\dagger \bu\leq (1-\alpha) P_r}\log
\left(1+\frac{|\sum_{m=1}^M z_m u_m|^2}{N_t}\right)\\
&=\log \left(1+\frac{\max_{\bu^\dagger \bu\leq (1-\alpha) P_r} (\bu^\dagger \z \z^\dagger \bu)}{N_t}\right)\\
&=\log \left(1+\frac{\max_{\bv^\dagger \bv\leq (1-\alpha) P_r}
(\bv^\dagger{\hh_{h}^\bot}^\dagger \z \z^\dagger
{\hh_{h}^\bot}\bv)}{N_t}\right)\\
&=\log \left(1+\frac{(1-\alpha)P_r\lambda_{max}
({\hh_{h}^\bot}^\dagger \z \z^\dagger {\hh_{h}^\bot})}{N_t}\right)\label{a}\\
&=\log \left(1+\frac{(1-\alpha)P_r \z^\dagger
{\hh_{h}^\bot}{\hh_{h}^\bot}^\dagger \z}{N_t}\right)\label{b}
\end{align}
The optimum beamforming vector $\bu$ is
\begin{align}\label{woptdfto}
\bu_{opt}=\hh_{h}^\bot \bv=\varsigma_1 \hh_{h}^\bot
{\hh_{h}^\bot}^\dagger \z
\end{align}
where $\varsigma_1$ is a constant introduced to satisfy the power constraint. Hence, secrecy rate region
$\mathbb{R}_{s,b}$ achieved with this strategy is
\begin{align}
\begin{split}
0\leq R_d \leq R_{d,m}(\h,\z,P_r,\alpha)\\
 0 \leq R_e \leq
R_{e,m}(\h,\z,P_r,\alpha)
\end{split}\label{sinb}
\end{align}

Note that we can switch the role of $D$ and $E$, and choose $\w$ to be in
the null space of $\z^\dagger$. In general, the union of region described in (\ref{sinb})
and its switched counterpart is the secrecy rate
region of single null space beamforming strategy.

\subsection{Double Null Space Beamforming}
In this scheme, we simultaneously choose the beamforming vectors for
$D$ and $E$ to lie in the null space of each other's channel vector.
That is $|\sum_{m=1}^M h_m u_m|^2=\h^\dagger \bu=0$,  and $|\sum_{m=1}^M
z_m w_m|^2=\z^\dagger \w=0$. In this case, the channel reduces to two
parallel channels. Since interference is completely eliminated,
the secrecy constraint is automatically satisfied. 
Coding for secrecy is not needed at the relays. The channel input-output relations are
\begin{align}
y_d&=\h^\dagger \w x_d  +n_0 \\
 y_e&=\mathbf{z}^\dagger  \bu x_e+n_1
\end{align}
Now, we only need to solve the following problems:
\begin{align}
\max_{\w^\dagger \w\leq \alpha P_r}\log \left(1+\frac{|\sum_{m=1}^M
h_m w_m|^2}{N_0}\right) ~~~~~s.t ~~\z^\dagger \w=0 \label{dbrd}\\
\max_{\bu^\dagger \bu\leq (1-\alpha) P_r}\log
\left(1+\frac{|\sum_{m=1}^M z_m u_m|^2}{N_0}\right) ~~~~~s.t
~~\h^\dagger \bu=0.
\end{align}
Similarly as in Section \ref{subsec:singlenull}, we can easily find the secrecy
rate region  $\mathbb{R}_{d,b}$ for double null space beamforming as
\begin{align}
&0 \leq R_d \leq \log \left(1+\frac{\alpha P_r \h^\dagger
{\hh_{z}^\bot}{\hh_{z}^\bot}^\dagger \h}{N_0}\right)\label{db1}\\
&0 \leq R_e \leq \log \left(1+\frac{(1-\alpha)P_r \z^\dagger
{\hh_{h}^\bot}{\hh_{h}^\bot}^\dagger \z}{N_0}\right)\label{db2}
\end{align}
where $\hh_{z}^\bot$ denote the projection matrix onto the null
space of $\z^\dagger$ and is defined similarly  as $\hh_{h}^\bot$.

\subsection{TDMA}
For comparison, we consider in the second-hop that the relay only transmits
secret information to one user at a time and treat the other user as
the eavesdropper. We assume that relay uses $\alpha$ fraction of time to
transmit $x_d$ where $(1-\alpha)$ fraction of the time is used to transmit
$x_e$. The channel now is the standard gaussian wiretap channel
instead of an interference channel. It can be easily shown that the rate
region $\mathbb{R}_{tdma}$ is
\begin{align}
0\leq &R_d\leq \alpha \log \lambda_{max}(N_0\I+P_r\h
\h^\dagger,N_0\I+P_r\z\z^\dagger)\\
0\leq &R_e \leq (1-\alpha)\log \lambda_{max}(N_0\I+P_r\z
\z^\dagger,N_0\I+P_r\h\h^\dagger)
\end{align}

\section{Optimality}
In this section, we investigate the optimality of our proposed null
space beamforming techniques. Although the optimal values of $\w$ and
$\bu$ that maximize the rate region (\ref{Rd}) and (\ref{Re}) is unknown,
we can easily see that the following rate region is an outer bound
region of our original achievable secrecy rate region.
\begin{align}\label{out}
0 \leq R_d \leq &\log \left(1+\frac{|\sum_{m=1}^M h_m
w_m|^2}{N_0}\right)
\\
&- \log \left( 1+\frac{|\sum_{m=1}^M z_m w_m|^2}{N_0}\right)
\\
0 \leq R_e \leq &\log \left(1+\frac{|\sum_{m=1}^M z_m
u_m|^2}{N_0}\right)
\\
&- \log \left( 1+\frac{|\sum_{m=1}^M h_m
u_m|^2}{N_0}\right).
\end{align}
Again, this rate region should be maximized with all possible $\w$
and $\bu$ satisfying $||\w||^2+||\bu||^2\leq P_r$. From the above
expressions, we can see that this outer bound can be interpreted as 
two simultaneously transmitting wire-tap channels. Fortunately, the optimization problem in this case can be solved analytically. With the
same assumptions as before that $||\w||^2=\alpha P_r$,
$||\bu||^2=(1-\alpha) P_r$, we can easily show that the outer bound
secrecy rate region $\mathbb{R}_{outer}$ of our collaborative relay
beamforming system is
\begin{align}
0\leq &R_d\leq  \log \lambda_{max}(N_0\I+\alpha P_r\h
\h^\dagger,N_0\I+\alpha P_r \z\z^\dagger)\label{rdout}\\
0\leq &R_e \leq \log \lambda_{max}(N_0\I+(1-\alpha)P_r\z
\z^\dagger,N_0\I+(1-\alpha)P_r\h\h^\dagger)\label{reout}
\end{align}
The expression for $R_d$ and $R_e$ here coincide with the secrecy
capacity of Gaussian MISO wiretap channel \cite{shafiee}
\cite{khisti} with transmit power levels $\alpha P$ and $(1-\alpha) P$.

%As we know, we can think
%of our $M$ relay as the single transmitter with $M$ transmitting
%antennas, The MISO secrecy broadcasting capacity region
%$\mathbb{R}_{MISO}$ is derived in \cite{Liu1}, which can be treated
%as the upper bound rate region of our collaborative relay
%beamforming system. Hence, we will use this rate region as the
%benchmark of our proposed scheme. As derived in \cite{Liu1}, the
%region can be written as
%\begin{align}\label{MISO}
%0\leq R_d \leq \log(\gamma_1(\alpha)) \\
%0\leq R_e \leq \log(\gamma_2(\alpha))
%\end{align}
%where
%\begin{align}
%&\gamma_1(\alpha)=\frac{1+\alpha P |\h^\dagger \e_1|^2}{1+\alpha P
%|\z^\dagger \e_1|^2} \\
%&\gamma_2(\alpha)=\lambda_{max}\left(\I+\frac{(1-\alpha)P \z
%\z^\dagger}{1+\alpha P |\z^\dagger \e_1|^2},\I+\frac{(1-\alpha)P \h
%\h^\dagger}{1+\alpha P |\h^\dagger \e_1|^2}\right)
%\end{align}
%\e_1$ denote the normalized eigenvector corresponding to largest
%eigenvalue of the pencil $(\I+P \h \h^ \dagger, I +P \z
%\z^\dagger)$. Note, that for simplicity we assume $N_0=1$, which
%will also be assumed in following discussion. and $P$ is the
%transmitter power, which is corresponding to $P_r$ in our
%collaborative relay setting.

\subsection{Optimality in the High-SNR Regime}
% We first note that
%\begin{align}
%\gamma_1(\alpha)&=\frac{1+\alpha P| \h^\dagger \e_1|^2}{1+\alpha P
%|\z^\dagger \e_1|^2}\\
%&=\frac{\e_1^\dagger(\I+\alpha P \h
%\h^\dagger)\e_1}{\e_1^\dagger(\I+\alpha P \z \z^\dagger)\e_1}\\
%&\leq \lambda_{max}(\I+\alpha P \h \h^\dagger,\I+\alpha P \z
%\z^\dagger)\label{gamma1}
%\end{align}
%
%\begin{align}
%\gamma_2(\alpha)&=\lambda_{max}\left(\I+\frac{(1-\alpha)P \z
%\z^\dagger}{1+\alpha P |\z^\dagger \e_1|^2},\I+\frac{(1-\alpha)P \h
%\h^\dagger}{1+\alpha P |\h^\dagger \e_1|^2}\right)\\
%&=max \frac{\psi^\dagger ((1+\alpha P |\h^\dagger
%\e_1|^2)\I+\frac{1+\alpha P |\h^\dagger \e_1|^2}{(1+\alpha P
%|\z^\dagger \e_1|^2}(1-\alpha)P \z \z^\dagger )\psi}{\psi^\dagger
%((1+\alpha P |\h^\dagger
%\e_1|^2)\I+(1-\alpha)P \h \h^\dagger)\psi}\\
%&=max \frac{\psi^\dagger \alpha P |\z^\dagger \e_1|^2 \I \psi
%+\psi^\dagger (\I+(1-\alpha)P \z \z^\dagger )\psi}{\psi^\dagger
%\alpha P |\h^\dagger \e_1|^2\I\psi +\psi^\dagger(\I+(1-\alpha)P \h
%\h^\dagger)\psi}\\
%&=max \frac{\alpha P |\z^\dagger \e_1|^2 +\psi^\dagger
%(\I+(1-\alpha)P \z \z^\dagger )\psi}{\alpha P |\h^\dagger
%\e_1|^2+\psi^\dagger(\I+(1-\alpha)P \h \h^\dagger)\psi}\\
%&\leq max \frac{\psi^\dagger (\I+(1-\alpha)P \z \z^\dagger
%)\psi}{\psi^\dagger(\I+(1-\alpha)P \h \h^\dagger)\psi}\label{dddd}\\
%&=\lambda_{max}(\I+(1-\alpha)P \z \z^\dagger,\I+(1-\alpha)P \h
%\h^\dagger)\label{gamma2}
%\end{align}
%where getting (\ref{dddd}) we use the fact that$|\h^\dagger
%\e_1|^2|>|\z^\dagger \e_1|^2$ and $max \frac{\psi^\dagger
%(\I+(1-\alpha)P \z \z^\dagger )\psi}{\psi^\dagger(\I+(1-\alpha)P \h
%\h^\dagger)\psi}>1$.

In this section, we show that the outer bound region $\mathbb{R}_{outer}$
converges to the proposed null space beamforming regions at high
SNR. For the single null space beamforming scheme,  the maximum $R_d$ in (\ref{ra}) has the same
express as in (\ref{rdout}), and thus it is automatically optimal. $R_e$ in single null space beamforming has
basically the same expression as that of $R_e$ in double null space
beamforming with $N_0$ replaced by $N_t$. This difference is
negligible as $P$ goes infinity. Hence, we focus on double null
space beamforming and show that in the high-SNR regime, the
$\mathbb{R}_{outer}$  coincide with the double null space region described by 
(\ref{db1}) and (\ref{db2}). In the following analysis, for simplicity and without loss of generality, we assume $N_0=1$.
From the
Corollary $4$ in Chapter $4$ of  \cite{khisti}, we can see that
\begin{align}
\lim_{P_r \to \infty}\frac{1}{P_r}\lambda_{max}(\I+ P_r \h
\h^\dagger, \I+P_r \z \z^\dagger) =\max_{\tilde{\psi}} |\h^\dagger
\tilde{\psi}|^2\label{highsnr}
\end{align}
where $\tilde{\psi}$ is a unit vector on the null space of $\z^\dagger$.
Similarly, we can define $\tilde{\psi_1}$ as a unit vector on the
null space of $\h^\dagger$. Combining this result with
(\ref{rdout}) and (\ref{reout}), we can express the region
$\mathbb{R}_{outer}$ at high SNRs as
\begin{align}
0\leq R_d \leq \log(\alpha P_r)+\log(\max_{\tilde{\psi}} |\h^\dagger
\tilde{\psi}|^2) + o(1) \\
0\leq R_e \leq \log((1-\alpha) P_r)+\log(\max_{\tilde{\psi_1}}
|\z^\dagger \tilde{\psi_1}|^2) + o(1)
\end{align}
where $o(1) \to 0$ as $P_r \to \infty$.
On the
other hand, double null space beamforming region satisfies
\begin{align}
0 \leq R_d &\leq \max_{\w^\dagger
\w\leq \alpha P_r}\log \left(1+|\sum_{m=1}^M h_m w_m|^2\right)\\
&= \log(\alpha P_r)+\log(\max_{\tilde{\psi}} |\h^\dagger
\tilde{\psi}|^2) + o(1) 
\label{eq:asymbound}\\
0 \leq  R_e &\leq 
\max_{\bu^\dagger
\bu\leq (1-\alpha) P_r}\log \left(1+|\sum_{m=1}^M z_m u_m|^2\right)\\
&= \log((1-\alpha) P_r)+\log(\max_{\tilde{\psi_1}} |\z^\dagger
\tilde{\psi_1}|^2) +o(1). \label{eq:asymbound2}
\end{align}
Above, (\ref{eq:asymbound}) follows from the observation that
\begin{align}
&\lim_{P_r \to \infty} \log \left(1+|\sum_{m=1}^M h_m w_m|^2\right) - \log(\alpha P_r)
\\
&= \lim_{P_r \to \infty} \log \left(\frac{1}{\alpha P_r}+\left|\sum_{m=1}^M h_m \frac{w_m}{\sqrt{\alpha P_r}}\right|^2\right)
\\
&=\log |\h^\dagger \tilde{\psi}|^2
\end{align}
where $\tilde{\psi}$ is a unit vector and is in the null space of $\z^\dagger$ because $\w$ is in the null space of $\z^\dagger$. (\ref{eq:asymbound2}) follows similarly.
Thus, the outer bound secrecy rate region converges to the double null
space beamforming region in the high-SNR regime, showing that  the null space
beamforming strategies are optimal in this regime.

\subsection{Optimality of TDMA in the Low-SNR Regime}
In this section, we consider the limit $P_r\to 0$.   In the following steps, the
order notation $o(P_r)$ means that $o(P_r)/P_r \rightarrow 0$ as
$P_r \rightarrow 0$.
\begin{align}
&\lambda_{max}(\I + P_r\h \h^\dagger, \I + P_r\z \z^\dagger)\\
&= \lambda_{max}\left((\I + P_r\z \z^\dagger)^{-1}(\I +
P_r\h\h^\dagger)\right)\\
&= \lambda_{max}\left((\I - P_r\z \z^\dagger  + o(P_r))(\I +
P_r\h\h^\dagger)\right)\\
&= \lambda_{max}\left((\I - P_r\z^\dagger \z )(\I +
P_r\h\h^\dagger)\right) + o(P_r) \\
&= \lambda_{max}\left(\I + P_r(\h\h^\dagger - \z\z^\dagger
)\right)+o(P_r)\\
&= 1 + P_r\lambda_{max}(\h \h^\dagger - \z\z^\dagger) + o(P_r)
\label{lowSNR}
\end{align}
Combining this low-SNR approximation with (\ref{rdout}) and (\ref{reout}), we can see that the
$\mathbb{R}_{outer}$ at low SNRs is {\small
\begin{align}
0\leq R_d &\leq \log\lambda_{max}(\I+\alpha
P_r \h \h^\dagger,\I+\alpha P_r \z \z^\dagger) \nonumber \\
&= \alpha P_r\lambda_{max}(\h \h^\dagger - \z\z^\dagger) + o(P_r)\label{lrd}\\
0 \leq R_e &\leq\log\lambda_{max}(\I+(1-\alpha)
P_r \z \z^\dagger,\I+(1-\alpha P_r) \h \h^\dagger) \nonumber \\
&= (1-\alpha )P_r\lambda_{max}(\z \z^\dagger -
\h\h^\dagger) + o(P_r)\label{lre}
\end{align}}
Note that (\ref{lrd}) and (\ref{lre}) are also the low-SNR approximations for
the TDMA approach. Thus, the TDMA scheme can achieve the optimal rate
region in the low-SNR regime. For the completeness, we give the lower SNR
approximations for single and double null space beamforming as well. For
single null space beamforming scheme, the low-SNR approximation of
(\ref{sinb}) is
\begin{align}
0\leq R_d \leq \alpha P_r\lambda_{max}(\h \h^\dagger - \z\z^\dagger) + o(P_r)\\
0\leq R_e \leq (1-\alpha) P_r/N_t \z^\dagger
{\hh_{z}^\bot}{\hh_{z}^\bot}^\dagger \z + o(P_r)
\end{align}
while for the double null space  beamforming scheme, low-SNR
approximations of (\ref{db1}) and (\ref{db2}) are
\begin{align}
0\leq R_d \leq \alpha P_r \h^\dagger
{\hh_{z}^\bot}{\hh_{z}^\bot}^\dagger \h + o(P_r)\\
0\leq R_e \leq (1-\alpha )P_r \z^\dagger
{\hh_{h}^\bot}{\hh_{h}^\bot}^\dagger \z + o(P_r)
\end{align}

\subsection{Optimality when the Number of Relays is Large}
It is easy to show that
\begin{align}
\lambda_{max}(\I+\alpha P_r \h
\h^\dagger,\I+\alpha P_r \z \z^\dagger) &\leq \lambda_{max}(\I+\alpha P_r \h \h^\dagger) \nonumber \\
&=1+\alpha P_r \h^\dagger \h \label{aaa}
\end{align}
Now, consider the function
\begin{align}
1+\alpha P_r \h^\dagger {\hh_{z}^\bot}{\hh_{z}^\bot}^\dagger
\h\label{bbb}
\end{align}
which is inside the $\log$ function in the double null space beamforming
$R_d$ boundary rate (\ref{db1}). In our numerical results, we observe that when $M$ is
large and $\h$ and $\z$ are Gaussian distributed (Rayleigh fading environment),
(\ref{aaa}) and (\ref{bbb}) converge to the same value. Similar results
are also noted when $R_e$ in (\ref{db2}) is considered. These numerical observations indicate the optimality of null space beamforming strategies in the regime in which the number of relays, $M$, is large.

\section{simulation results}

In our simulations, we assume $N_m=N_0=1$, and $\{g_m\}$, $\{h_m\},
\{z_m\}$ are complex, circularly symmetric Gaussian random variables
with zero mean and variances $\sigma_g^2$, $\sigma_h^2$, and
$\sigma_z^2$ respectively.

\begin{figure}
\begin{center}
\includegraphics[width = 0.4\textwidth]{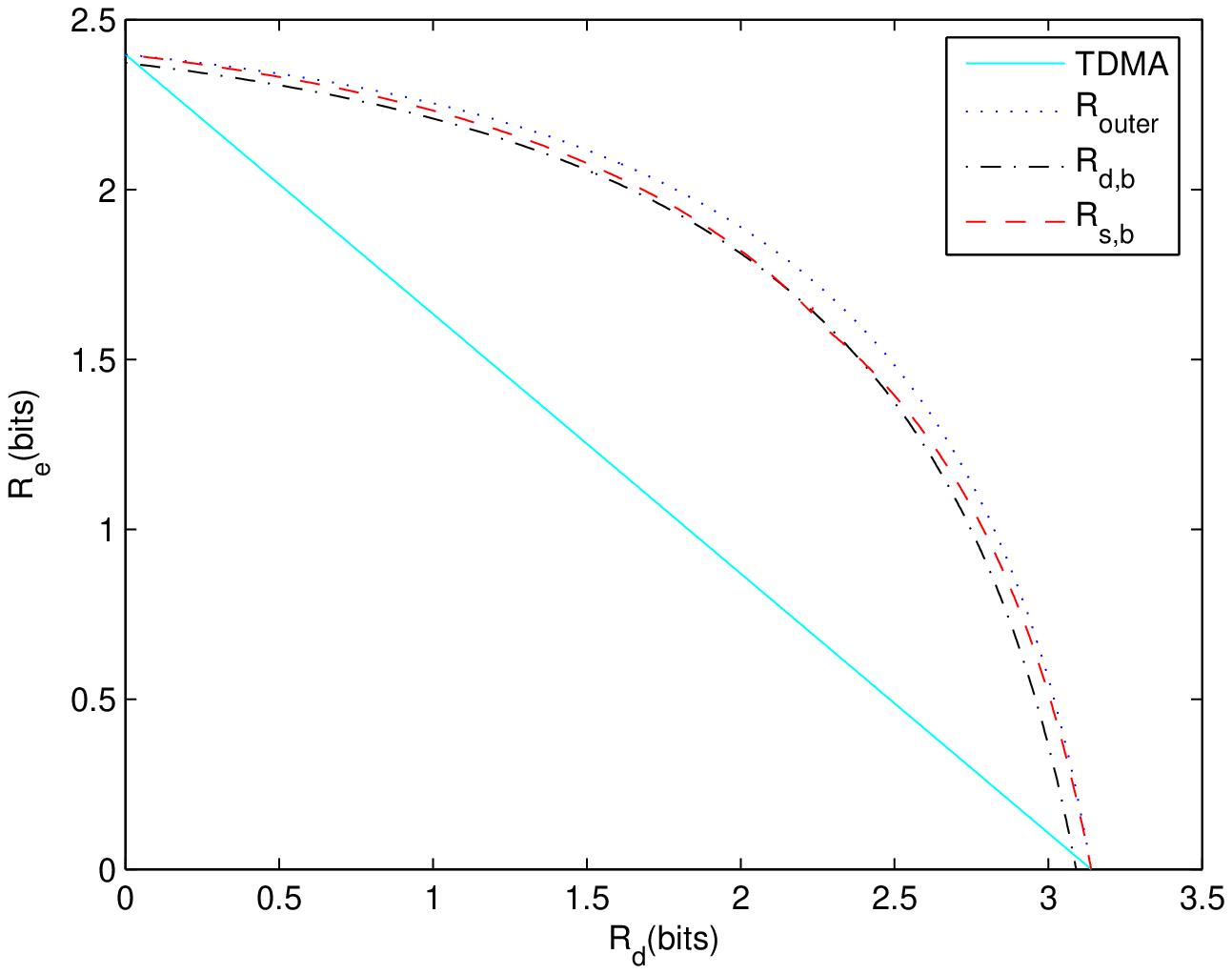}
\includegraphics[width = 0.4\textwidth]{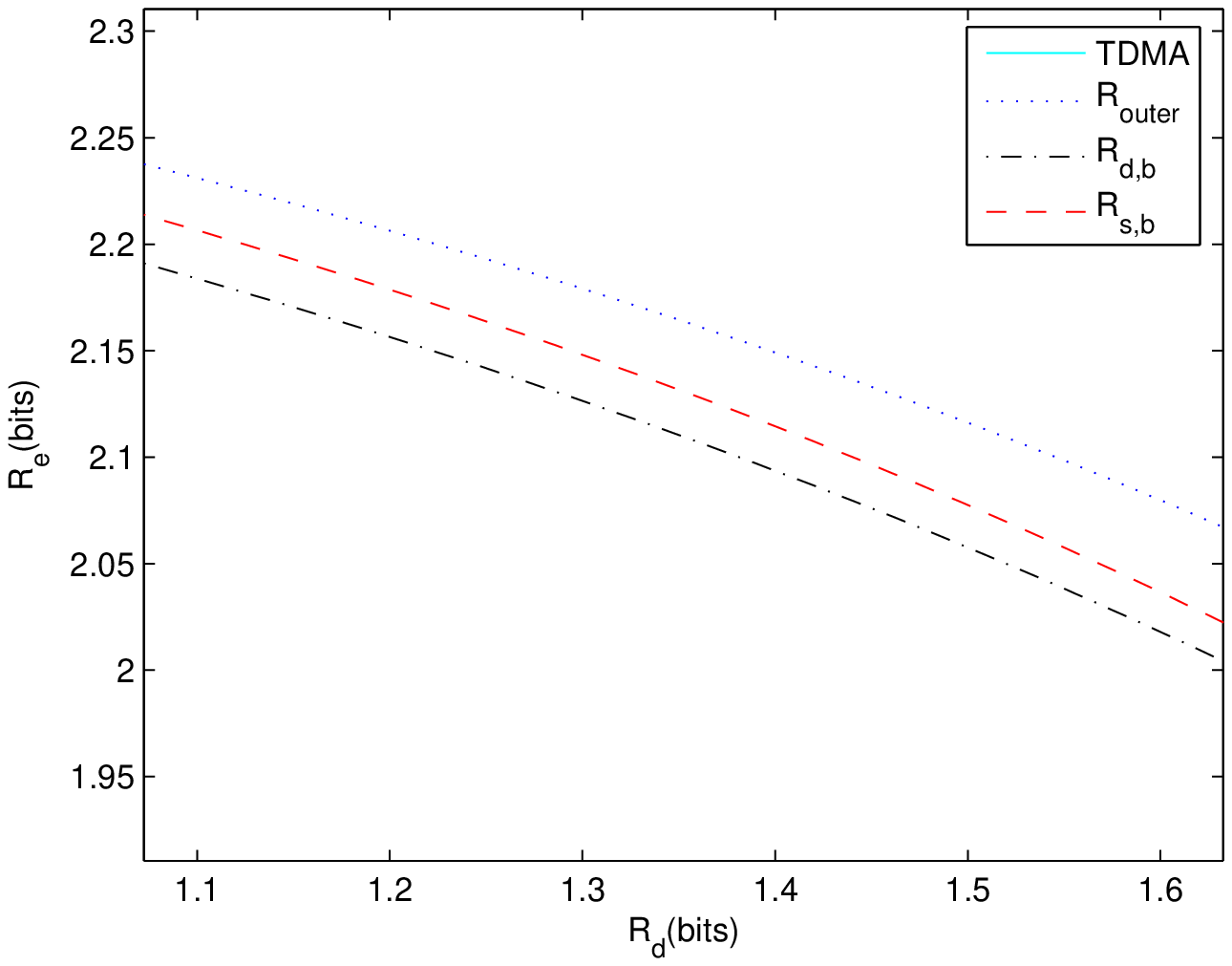}
\caption{Second-hop secrecy rate region $  \sigma_h=2,
\sigma_z=2, P_r=1, M=5$. Lower figure provides a zoomed version.} \label{fig:region1}
\end{center}
\end{figure}

\begin{figure}
\begin{center}
\includegraphics[width = 0.4\textwidth]{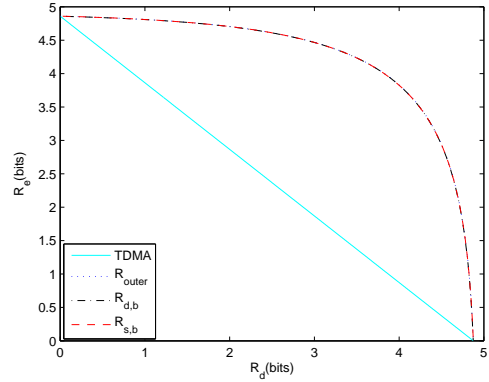}
\caption{Second-hop secrecy rate region $ \sigma_h=2,
\sigma_z=2, P_r=1, M=15$ } \label{fig:region2}
\end{center}
\end{figure}

In Figures \ref{fig:region1} and \ref{fig:region2}, we plot the
second-hop secrecy rate region of different schemes in which we see
$\mathbb{R}_{outer}\supset\mathbb{R}_{s,b}\supset
\mathbb{R}_{d,b}\supset \mathbb{R}_{tdma}$. We notice that our
proposed suboptimal beamforming region is very close to outer bound
secrecy region $\mathbb{R}_{outer}$. Furthermore, the larger the
$M$, the smaller the rate gap between $\mathbb{R}_{outer}$ and our
proposed null space beamforming schemes. Also, we note that increasing the number of relays,$M$, enlargens the rate region. Moreover, we can see that $M=15$ is
sufficient for the null space beamforming schemes to coincide with
the $\mathbb{R}_{outer}$.

\begin{figure}
\begin{center}
\includegraphics[width = 0.4\textwidth]{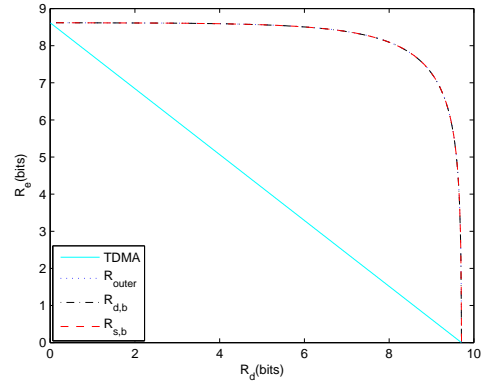}
\caption{Second hop secrecy rate region $  \sigma_h=2,
\sigma_z=2, P_r=100,  M=3$ } \label{fig:region4}
\end{center}
\end{figure}

\begin{figure}
\begin{center}
\includegraphics[width = 0.4\textwidth]{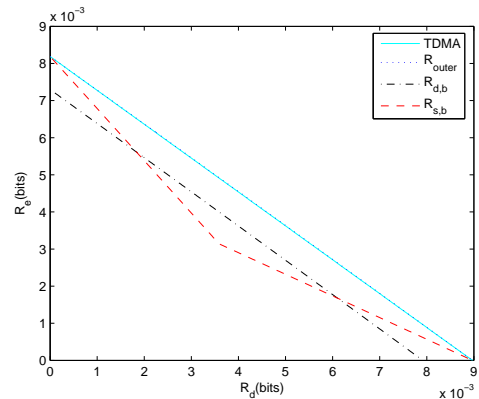}
\caption{Second hop secrecy rate region $ \sigma_h=2,
\sigma_z=2, P_r=0.001,  M=10$ } \label{fig:regionlow}
\end{center}
\end{figure}

Next, we examine the null space beamforming's optimality in the high-SNR
regime in Fig. \ref{fig:region4}. In this simulation, we can see that when the
relay power is large enough, $\mathbb{R}_{outer}$ coincides
with the regions of our proposed null space beamforming schemes as expected even
$M$ is very small. Finally, in Fig. \ref{fig:regionlow} where relay
power small, we observe that $\mathbb{R}_{outer}$ coincides with the rate region of the
TDMA transmission scheme. Also, we note  that the double null space
beamforming has better performance than single null space
beamforming at some operation points. This is mainly because $N_t$
is no longer negligible at very low SNR values.

\section{Conclusion}

In this paper, we have considered a DF-based collaborative relay
beamforming protocol to achieve secure broadcasting to two users. As
the general optimization of relay weights is a difficult task, we have proposed
single and double null space beamforming schemes. We have compared the rate regions of these 
two schemes and the TDMA scheme with the outer bound secrecy rate
region of the original the relay beamforming system.  We have analytically shown that null space beamforming
schemes are optimal in the high-SNR regime, and TDMA scheme is optimal
in the low-SNR regime. In our numerical results, we have seen that our
proposed null space beamforming schemes perform in general very close to outer bound secrecy rate region. We have numerically shown that when the number of
relays is large, the null space beamforming schemes are optimal.

\end{document}